\documentclass[runningheads]{llncs}
\usepackage{amsmath}
\usepackage{multirow}
\usepackage{booktabs}
\usepackage{graphicx}
\usepackage[table,xcdraw]{xcolor}
\usepackage[table,xcdraw]{xcolor}
\begin{document}
\title{Strengthening Network Intrusion Detection in IoT Environments with Self-Supervised Learning and Few Shot Learning}
\titlerunning{Strengthening Network Intrusion Detection}
%

\author{Safa Ben Atitallah\inst{1,2}
\and
Maha Driss\inst{1,2} 
\and
Wadii Boulila\inst{1,2}
\and 
Anis Koubaa\inst{1}
}
%
\institute{ RIOTU Lab, CCIS, Prince Sultan University, Riyadh 12435, Saudi Arabia \and RIADI Laboratory, University of Manouba, Manouba 2010, Tunisia
}
\maketitle              
\begin{abstract}
The Internet of Things (IoT) has been introduced as a breakthrough technology that integrates intelligence into everyday objects, enabling high levels of connectivity between them. As the IoT networks grow and expand, they become more susceptible to cybersecurity attacks.
A significant challenge in current intrusion detection systems for IoT includes handling imbalanced datasets where labeled data are scarce, particularly for new and rare types of cyber attacks. Existing literature often fails to detect such underrepresented attack classes.
This paper introduces a novel intrusion detection approach designed to address these challenges. By integrating Self Supervised Learning (SSL), Few Shot Learning (FSL), and Random Forest (RF), our approach excels in learning from limited and imbalanced data and enhancing detection capabilities. The approach starts with a Deep Infomax model trained to extract key features from the dataset. These features are then fed into a prototypical network to generate discriminate embedding. Subsequently, an RF classifier is employed to detect and classify potential malware, including a range of attacks that are frequently observed in IoT networks. The proposed approach was evaluated through two different datasets, MaleVis and WSN-DS, which demonstrate its superior performance 
with accuracies of 98.60\% and 99.56\%, precisions of 98.79\% and 99.56\%, recalls of 98.60\% and 99.56\%, and F1-scores of 98.63\% and 99.56\%, respectively.
\end{abstract}
\section{Introduction}
Recently, the fast development of intrusion traffic has posed several cybersecurity challenges \cite{atitallah2020leveraging}. Traditional detection algorithms faced a significant limitation related to the shortage of labeled data for new and unusual attack types, which leads to imbalanced datasets \cite{lee2020internet}. This challenge is mainly prominent in Internet of Things (IoT) environments, where the dynamic nature of linked devices gives rise to new attack patterns that are not well-represented in labeled datasets \cite{yu2020intrusion}. This lack of labeled data severely impacts the performance of supervised learning algorithms, which mostly rely on big labeled data to detect and effectively block such attacks. In response to this issue, our work presents a novel intrusion detection approach that combines the advantages of Few Shot Learning (FSL) and Self-Supervised Learning (SSL). 
Our proposed approach leverages SSL, particularly the Deep InfoMax (DIM). DIM, a contrastive learning model, is an excellent tool for identifying unusual patterns and abnormalities in network traffic which help to identify potential intrusions. This model is trained to learn how to extract comprehensive and relevant features from unlabelled data in cybersecurity scenarios.
Furthermore, we employ the FSL techniques to enhance the model's ability to detect new and uncommon intrusion attacks using only limited labeled samples. We used prototypical networks, which are well known for their strong ability to identify similarities, to detect minor distinctions between known attack patterns and new threats. 
After obtaining the embedding output from the ProtoNet, we further classify it using the Random Forest (RF) algorithm, a powerful ensemble learning method.
This combination of SSL for feature extraction, FSL for discriminate embedding generation, and RF for classification form a robust approach that significantly improves the detection of novel intrusions. This approach is efficient in learning from imbalanced and limited labeled data and effective in recognizing emerging threats.
\textcolor{black}{The evaluation of this approach was conducted by using 2 datasets including MaleVis \cite{MaleVis} and WSN-DS \cite{almomani2016wsn} datasets. These datasets leverage a visualization strategy, converting attack records into images, to enhance pattern recognition and classification. }
Our novel approach for detecting and classifying intrusions in IoT environments offers several key contributions, which are summarized as follows:
\begin{itemize} 
    \item Showing exceptional learning efficiency with a small number of labeled samples, overcoming the challenges of relying on large annotated datasets.
    \item Significantly improving feature representation of samples by extracting comprehensive and relevant features 
    using the DIM model.
    \item  Ensuring robustness in the development of informative prototypes within prototypical networks.
    \item Enhancing the efficiency of intrusion detection systems in IoT networks, as demonstrated by the analysis of two malware datasets.
\end{itemize}

The rest of the paper is organized as follows. Section 2 provides a review of the related works that proposed intrusion detection systems for IoT environments, highlighting their main limitations. Section 3 presents the architecture of the proposed approach and describes its phases. Section 4 details and discusses the experiments conducted to evaluate the approach's performance. Finally, the paper concludes in Section 5, where suggestions about potential future research directions are outlined.
 
\section{Related Work}
Most of the current work in intrusion detection has been based on classical Machine/Deep Learning (ML/DL) techniques, which usually require large labeled and balanced datasets \cite{ullah2023tnn,latif2024dtl}. 
In \cite{ben2022effective}, the authors proposed a novel approach for detecting and classifying Denial-of-Service (DoS) attacks in Wireless Sensor Networks (WSNs) that employs pre-trained Convolutional Neural Networks (CNNs) and a majority voting mechanism. The proposed approach was tested on a dataset that included benign samples and four classes of DoS attacks, and it achieved good performance in detecting and categorizing the various classes of DoS attacks with high precision. However, the reliance on pre-trained CNNs may limit the ability of the model to generalize to novel attacks that were not present in the training phase.
The work presented in \cite{atitallah2022novel} proposed a novel approach for identifying and classifying IoT malware that employs a vision-based approach with deep transfer learning. It helped to improve the performance by combining ResNet18, MobileNetV2, and DenseNet161 using an RF voting technique rather than training models from scratch. 
This approach, tested on the MaleVis dataset, achieved a good performance in terms of precision, recall, specificity, F1-score, and accuracy while maintaining an acceptable processing time for classification. 
However, while this method reduces the need for training from scratch, it may suffer from overfitting and can not adapt to new malware types due to the static nature of transfer learning models.
The study in \cite{atitallah2023fedmicro} developed a new methodology that used a microservices-based architecture to divide IoT applications into tiny, autonomous, and interchangeable components. It employed federated learning to boost these microservices with intelligent data analysis capabilities, shift computing processes closer to the data source, minimize latency and network traffic, and improve data privacy. This approach was validated using an IoT malware detection scenario. 
However, federated learning can introduce complexities in managing data distribution and model convergence, affecting the system's scalability and efficiency.
\\
Recently, research in intrusion detection has explored innovative approaches to address the issues related to the limited availability of extensively labeled datasets, particularly concerning emerging attack types. 
In \cite{ayesha2023fs3}, Ayesha et al. introduced a novel framework for IoT network intrusion detection named FS3 to tackle the class imbalance in datasets. FS3 leverages the tabular multilayer perceptron (TabMLP) model to extract patterns from unlabelled data and then learn from a few labeled examples to detect and classify potential attacks effectively. The framework performed well by using only 20\% of the labeled data for training. The findings demonstrated notable gains in recall and precision based on tests using three IoT datasets. While this approach works with unbalanced datasets, its effectiveness may be compromised in highly dynamic environments where attack patterns frequently change.\\
The previously discussed studies provided different techniques for improving security and data processing in IoT environments, each with distinctive strengths and advantages. However, they also presented numerous limitations.
These include the requirement of large labeled training datasets, potential susceptibility to emerging attack types, and trade-offs between high performance and processing resource requirements. Additionally, these models often struggle to adapt to the dynamic, real-world conditions of IoT scenarios, which can differ from those present during model training and testing.


\section{Background}
This section provides a theoretical overview of SSL and FSL, focusing on their effectiveness in detecting and classifying malware.

\subsection{Self Supervised Learning }
Self-supervised learning, a recent innovation in ML, has been introduced as a novel technique that falls between supervised and unsupervised learning. SSL aims to learn useful representations from the data without needing human-provided labels \cite{liu2021self}. This model is particularly effective when labeled data is rare or difficult to collect or when unlabelled data is unavailable. SSL methods are categorized into generative and contrastive types \cite{ericsson2022self}. Generative SSL, including Generative Adversarial Networks (GANs) and Variational Autoencoders (VAEs), 
focuses on generating data by training an encoder to encode input into a vector and a decoder to reconstruct the input from this vector. Contrastive SSL, on the other hand, involves training an encoder to convert data into vectors to compare and assess similarity, with methods like Bootstrap Your Own Latent (BYOL), Swapping Assignments between Views (SwAV), and Deep InfoMax (DIM) \cite{jaiswal2020survey}.
Our study leverages SSL, specifically the DIM model, in intrusion detection to extract usable representations from unlabelled data, therefore enhancing the accuracy of detecting various attack types.
DIM has been proposed as a DL approach that aims to maximize the mutual information between an input and and its encoded representation \cite{hjelm2018learning}. 
Through training, the DIM utilizes a contrastive learning approach, where the encoder is trained to distinguish between positive and negative pairs of samples. By concentrating on mutual information, the DIM can learn rich illustrations that include pertinent details about the data distribution.

\subsection{Few Sot Learning}
Recently, FSL has emerged as a powerful approach in ML \cite{wang2020generalizing}. It is based on recognizing and classifying samples through very few examples used in the training phase. Among the FSL, metric-based learning has been proposed, which involves creating a metric space where similar samples are near each other, facilitating their classification. This concept underlies models like Siamese, matching, and prototypical networks. The samples' classification is based on their proximity to the average prototype of each class. 
\\ In our work, we used the prototypical network to group similar attack types together for better classification. 
The idea of a prototypical network is simpler than the other AI models. It represents each class through its samples, called a prototype, in a representation space extracted through a specific neural network \cite{xiao2021adaptive}. The following equation \ref{eq:prototype} illustrates the calculation of the prototypes.
\begin{equation}
    cp_{i} = \frac{1}{K} \sum_{j=1}^K f_{\theta_{1}}(x^s_{ij})
    \label{eq:prototype}
\end{equation}
where $cp$ is the class prototype, $x^s_{ij}$ represents the embedding of the samples in the support set, and $k$ defines the total number of instances in each class.
\\
After determining the class prototypes with the support samples, a query sample is categorized by evaluating its distance to each prototype and attributing it to the class of the closest prototype \cite{snell2017prototypical}. 
The output probabilities assigned to each class are determined by applying a softmax function to the negative distances calculated, as demonstrated in the equation \ref{eq:2}.
\begin{equation}
    P(y=c|x) = softmax(-dist(f_{\theta_{1}}(x), cp_{i})) 
    \label{eq:2}
\end{equation}
After that, the loss function $L$ is defined as the negative natural logarithm of the likelihood associated with the correct class, as expressed in Equation \ref{eq:3}.
\begin{equation}
    L = - log P_{\theta_{1}} (y = c|x)
    \label{eq:3}
\end{equation}

\section{Proposed Approach}

\begin{figure}[ht]
    \centering
    \includegraphics[width=0.9\textwidth]{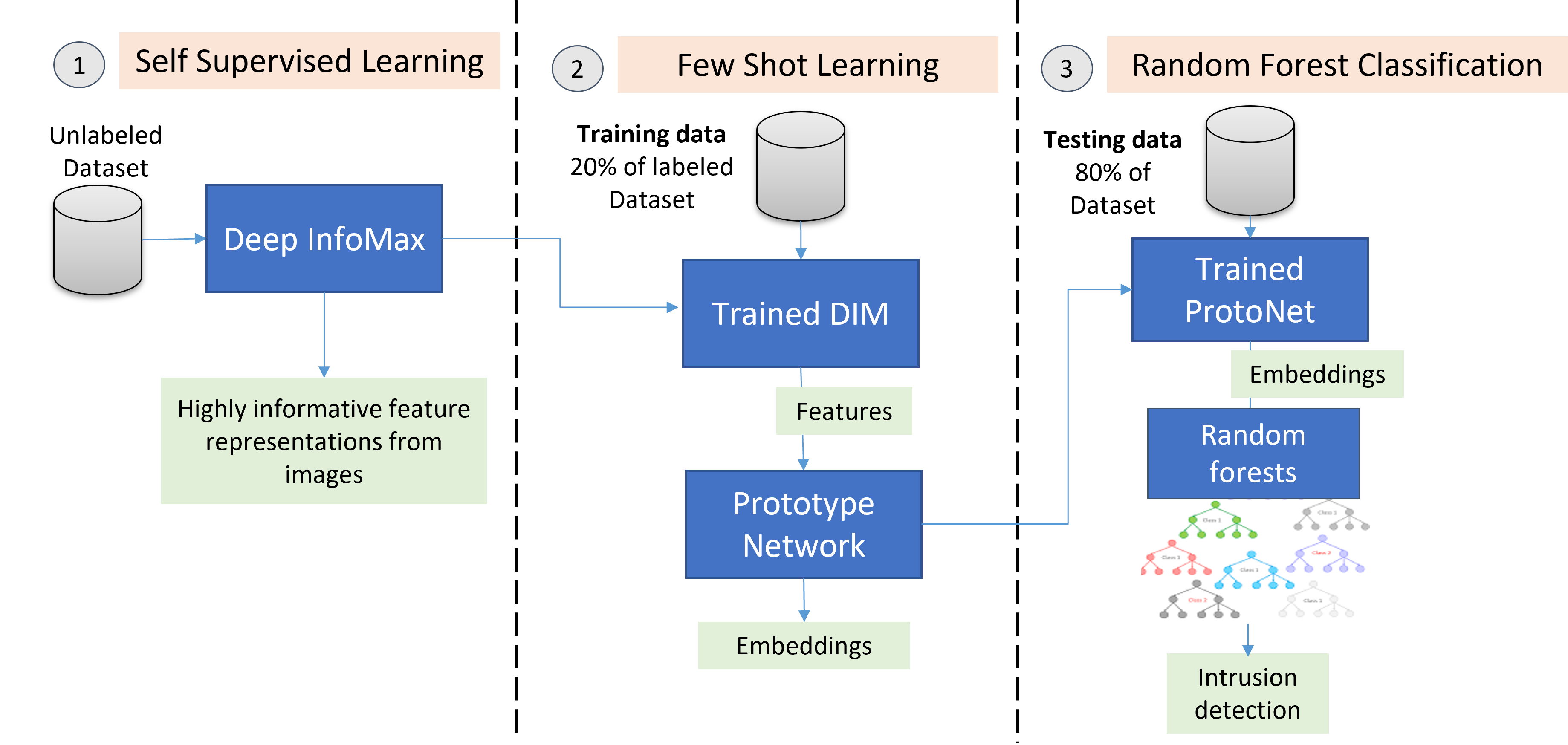}
    \caption{General overview of the proposed approach integrating SSL, FSL, and RF for intrusion detection classification.}
    \label{fig:approach}
\end{figure}

Our main objective in this study is to address the challenge of detecting new and emerging attacks, which often have significantly fewer labeled instances than other intrusion detection dataset classes. The process of collecting labeled data for each type of attack is both time-consuming and costly. 
Therefore, we aim to develop an approach that effectively detects these new attacks while operating without only a few labeled training examples.
\\
We propose adopting an FSL approach to address this challenge, particularly in intrusion detection datasets, explicitly utilizing the Prototypical Network (ProtoNet) framework \cite{snell2017prototypical}. 
ProtoNet operates by extracting embedding from image datasets to form class-specific prototypes. It then computes the distance between each query sample and these prototypes, assigning the sample to the class of the nearest prototype. This method is highly effective for FSL but can be further enhanced to achieve greater robustness and accuracy.
We suggest further improving our approach by integrating a self-supervised learning model within the ProtoNet to enhance its embedding extraction capabilities. For this purpose, we have selected the Deep InfoMax (DIM) model \cite{hjelm2018learning}, renowned for its ability to learn rich data representations by maximizing mutual information. This model's strength lies in its capacity to uncover intricate patterns and relationships within the data, which might be overlooked in traditional supervised learning methods.
By incorporating DIM into the ProtoNet framework, we aim to leverage its SSL power to extract more meaningful features from input images. These enhanced features will contribute to forming more informative and distinct prototypes for each class. The created prototypes are expected to encapsulate a deeper understanding of the data characteristics, therefore generating discriminate embedding. This embedding is further analyzed using an RF model, improving the classification accuracy of different attacks. 
For addressing the class imbalance in the intrusion dataset, our proposed approach, named DIM-ProtonetRF, is structured into three interconnected phases as depicted in Fig. \ref{fig:approach}, each leveraging the strengths of different ML models. In the following subsections, we will describe each phase of the proposed approach.
\vspace{-0.5cm}
\vspace{-1cm}

\subsection{Phase1: Self-Supervised Training of Deep InfoMax} 
The initial phase includes training the DIM model through the SSL setting. SSL is an effective technique that allows the model to learn and extract powerful and complex representations without using labeled data. During this phase, the DIM model captures the inherent patterns and structures found in the dataset by optimizing the mutual information between inputs and the corresponding high-level features.

\subsection{Phase2: Integrating DIM with Prototypical Network}
\begin{figure}[h]
    \centering
    \includegraphics[width=0.9\textwidth]{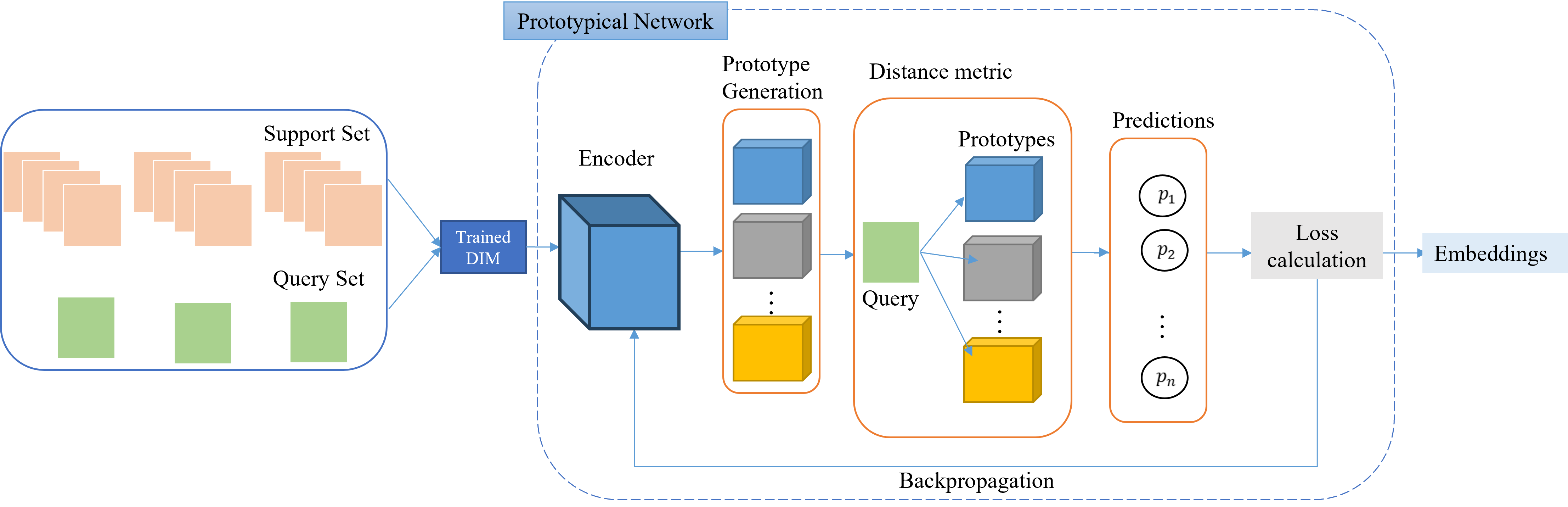}
    \caption{The architecture of the proposed DIM-based ProtoNet}
    \label{fig:fsl_id}
\end{figure}

In the next step of the proposed approach, we integrate SSL with FSL. This phase includes training the Prototypical Network, incorporating the previously trained DIM model. The proposed DIM-based ProtoNet is illustrated in Fig. \ref{fig:fsl_id}.
The integration of DIM enables ProtoNet to employ the enhanced and rich feature extraction capabilities developed during the first phase. The training now employs a limited set of labeled data, which aligns with the principles of the FSL paradigm. The ProtoNet, augmented with the power of DIM, extracts meaningful embedding from these few labeled samples, creating more robust and informative class prototypes.
As a result, the network forms more reliable and insightful prototypes for each class. 

\subsection{Phase 3: Classification with Random Forest}
In the final phase, we employ the RF model to classify the output embedding produced by the DIM-based  ProtoNet. The choice of RF is strategic, as it is known for its effectiveness in handling high-dimensional data and embedding. RF is an ensemble classifier that integrates multiple decision trees to improve accuracy and efficiency. It is particularly suitable for handling the embedding output from the DIM-based ProtoNet.
\\
This three-phase approach takes advantage of SSL, FSL, and ensemble methods. The integration between DIM's ability to learn rich features from unlabelled data, ProtoNet strength in handling few-shot scenarios, and RF robustness in classification tasks forms a powerful solution for tackling the challenges posed by class imbalance presented in malware datasets. Additionally, this combination enhances the approach's ability to detect new and emerging types of attacks, which increases its effectiveness in dynamic security environments.

\section{Experiments}
This section provides an overview of the experiment setup, describes the used datasets, and analyzes the outcomes of the proposed approach applied to two visual-based IoT malware datasets.

\subsection{Experimental Setup}
The experimental setup for the proposed method was conducted on a computer equipped with an 11th Generation Intel(R) Core(TM) i9-11900H processor, operating at a frequency of 2.50GHz with 32 GB of RAM, operating under Windows 11. For the development and coding of the approach, the Jupyter Notebook environment, part of the Anaconda, was utilized with Python version 3.8. 

\subsection{Datasets}

Our proposed intrusion detection system approach was evaluated using MaleVis and WSN-DS datasets. These datasets are designed to address specific intrusion and malware detection aspects within IoT environments. Below is a brief description of each dataset: 
\\
\textbf{Malevis dataset:}
This dataset is designed explicitly for malware visualization analysis \cite{MaleVis}. It offers a comprehensive collection of data points representing various malware signatures and patterns. This dataset consists of 25 distinct malware types alongside a benign class for comparative analysis. The benign class consists of many samples compared to the attack classes. 
\\
\textbf{WSN-DS dataset:}
The Wireless Sensor Network Dataset (WSN-DS) is collected to address security challenges in IoT wireless sensor networks \cite{almomani2016wsn}. It includes diverse data related to network traffic and sensor activities, providing a rich data source for experimentation. 
The WSN-DS contains four different classes of Denial of Service (DoS) attacks, including Blackhole, Grayhole, Flooding, Scheduling, and normal instances with an imbalanced distribution. 
The dataset, as modified in \cite{ben2022effective}, was utilized in our experiments, transforming from its original tabular format into images through the application of the Image Generator for Tabular Data (IGTD) algorithm.  

\subsection{Performance Evaluation}

For performance evaluation, a set of metrics is used, including: accuracy, precision, recall, F1-score, and confusion matrix.
These metrics assess the efficiency of intrusion detection systems and offer a comprehensive view of the overall performance of the detection system.
We train the DIM model on unlabelled data to learn how to extract rich representations from input data. Then, the trained DIM is used within the ProtoNet to help create more robust prototypes for each class in the dataset based on the selected support set. For each class, ten samples are selected for the support set. In the ProtoNet training, only 20\% of the data was used. The remaining set, which represents 80\%, was employed in the testing phase for detection and classification. 
The results of the DIM-ProtoNetRF model using the MaleVis dataset are depicted in Table \ref{tab:d1}. A reasonable accuracy rate of approximately 99\% was reached. Most classes are correctly detected and classified with an F1-score of 100\%.  
\begin{table}[h]
\centering
\caption{Performance results of the proposed DIM-ProtonetRF using the MaleVis dataset}
\label{tab:d1}
\begin{tabular}{|c|ccc|}
\hline
\textbf{Class}        &  \multicolumn{1}{c|}{\textbf{Precision}}             & \multicolumn{1}{c|}{\textbf{Recall}}                & \multicolumn{1}{c|}{\textbf{F1-score}  }            \\ \hline
Adposhel             & \multicolumn{1}{c|}{100}                         & \multicolumn{1}{c|}{100}                         & \multicolumn{1}{c|}{100}                            \\ \hline
Agent                 & \multicolumn{1}{c|}{100}                         & \multicolumn{1}{c|}{96.30}                         & \multicolumn{1}{c|}{98.11}                         \\ \hline
Allaple               &  \multicolumn{1}{c|}{100}                         & \multicolumn{1}{c|}{100}                         & \multicolumn{1}{c|}{100}                            \\ \hline
Amonetize             & \multicolumn{1}{c|}{100}                         & \multicolumn{1}{c|}{97.06}                         & \multicolumn{1}{c|}{98.51}                         \\ \hline
Androm                & \multicolumn{1}{c|}{100}                         & \multicolumn{1}{c|}{100}                         & \multicolumn{1}{c|}{100}                            \\ \hline
Autorun               & \multicolumn{1}{c|}{97.30}                         & \multicolumn{1}{c|}{97.30}                         & \multicolumn{1}{c|}{97.30  }                       \\ \hline
BrowseFox             &  \multicolumn{1}{c|}{100}                         & \multicolumn{1}{c|}{100}                         & \multicolumn{1}{c|}{100}                            \\ \hline
Dinwod                & \multicolumn{1}{c|}{100}                         & \multicolumn{1}{c|}{100}                         & \multicolumn{1}{c|}{100}                            \\ \hline
Elex                   & \multicolumn{1}{c|}{100}                         & \multicolumn{1}{c|}{100}                         & \multicolumn{1}{c|}{100}                            \\ \hline
Expiro                &\multicolumn{1}{c|}{93.75}                         & \multicolumn{1}{c|}{96.77}                         & \multicolumn{1}{c|}{95.24}                         \\ \hline
Fasong                & \multicolumn{1}{c|}{100}                         & \multicolumn{1}{c|}{100}                         & \multicolumn{1}{c|}{100}                            \\ \hline
HackKMS                & \multicolumn{1}{c|}{100}                         & \multicolumn{1}{c|}{100}                         & \multicolumn{1}{c|}{100}                            \\ \hline
Hlux                    & \multicolumn{1}{c|}{100}                         & \multicolumn{1}{c|}{100}                         & \multicolumn{1}{c|}{100}                            \\ \hline
Injector              & \multicolumn{1}{c|}{100}                         & \multicolumn{1}{c|}{96.43}                         & \multicolumn{1}{c|}{98.18}                         \\ \hline
InstallCore            & \multicolumn{1}{c|}{100}                         & \multicolumn{1}{c|}{100}                         & \multicolumn{1}{c|}{100}                            \\ \hline
MultiPlug             & \multicolumn{1}{c|}{96.55}                         & \multicolumn{1}{c|}{100}                         & \multicolumn{1}{c|}{98.25}                         \\ \hline
Neoreklami             & \multicolumn{1}{c|}{100}                         & \multicolumn{1}{c|}{100}                         & \multicolumn{1}{c|}{100}                            \\ \hline
Neshta                & \multicolumn{1}{c|}{95.24}                         & \multicolumn{1}{c|}{100}                         & \multicolumn{1}{c|}{97.56}                         \\ \hline
Other                 & \multicolumn{1}{c|}{83.91}                         & \multicolumn{1}{c|}{94.44}                         & \multicolumn{1}{c|}{82.93}                         \\ \hline
Regrun               & \multicolumn{1}{c|}{100}                         & \multicolumn{1}{c|}{100}                         & \multicolumn{1}{c|}{100}                            \\ \hline
Sality                & \multicolumn{1}{c|}{100}                         & \multicolumn{1}{c|}{83.87}                         & \multicolumn{1}{c|}{91.23}                         \\ \hline
Snarasite             & \multicolumn{1}{c|}{100}                         & \multicolumn{1}{c|}{100}                         & \multicolumn{1}{c|}{100}                            \\ \hline
Stantinko              & \multicolumn{1}{c|}{100}                         & \multicolumn{1}{c|}{100}                         & \multicolumn{1}{c|}{100}                            \\ \hline
VBA                   & \multicolumn{1}{c|}{100}                         & \multicolumn{1}{c|}{100}                         & \multicolumn{1}{c|}{100}                            \\ \hline
VBKryipt               & \multicolumn{1}{c|}{100}                         & \multicolumn{1}{c|}{100}                         & \multicolumn{1}{c|}{100}                            \\ \hline
Vilsel & \multicolumn{1}{c|}{100}                         & \multicolumn{1}{c|}{100}                         & \multicolumn{1}{c|}{100}                            \\ \hline
\rowcolor[HTML]{EFEFEF} 
\textbf{Accuracy}  & \multicolumn{3}{c|}{\cellcolor[HTML]{EFEFEF}98.60}  \\ \hline              \end{tabular}
\end{table}

\begin{figure}[h]
    \centering    \includegraphics[width=0.8\textwidth]{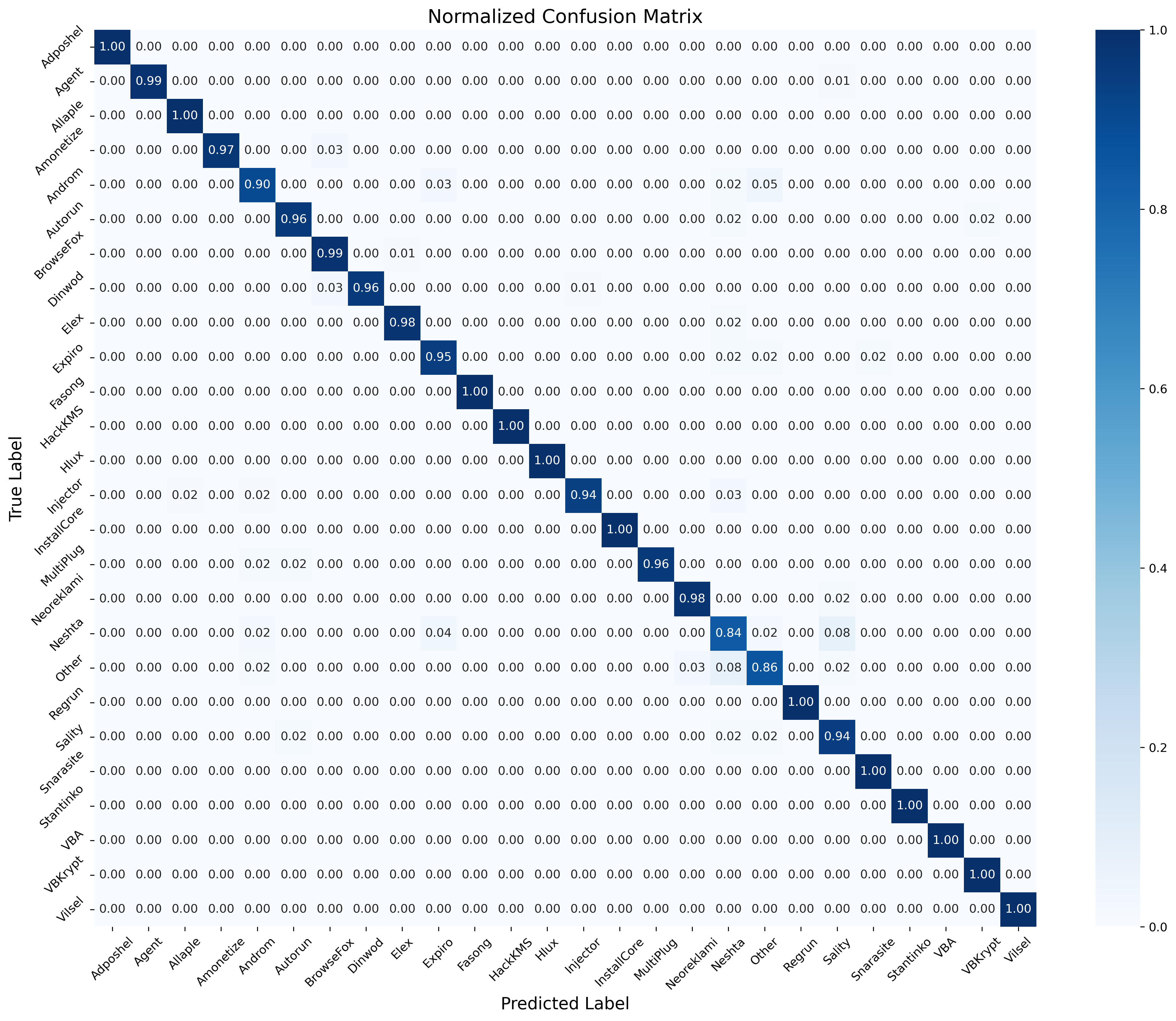} 
    \caption{Normalized confusion matrix within the MaleVis dataset following the proposed approach}
    \label{fig:m_CM}
\end{figure}

\begin{figure}[ht]
    \centering    \includegraphics[width=0.9\textwidth]{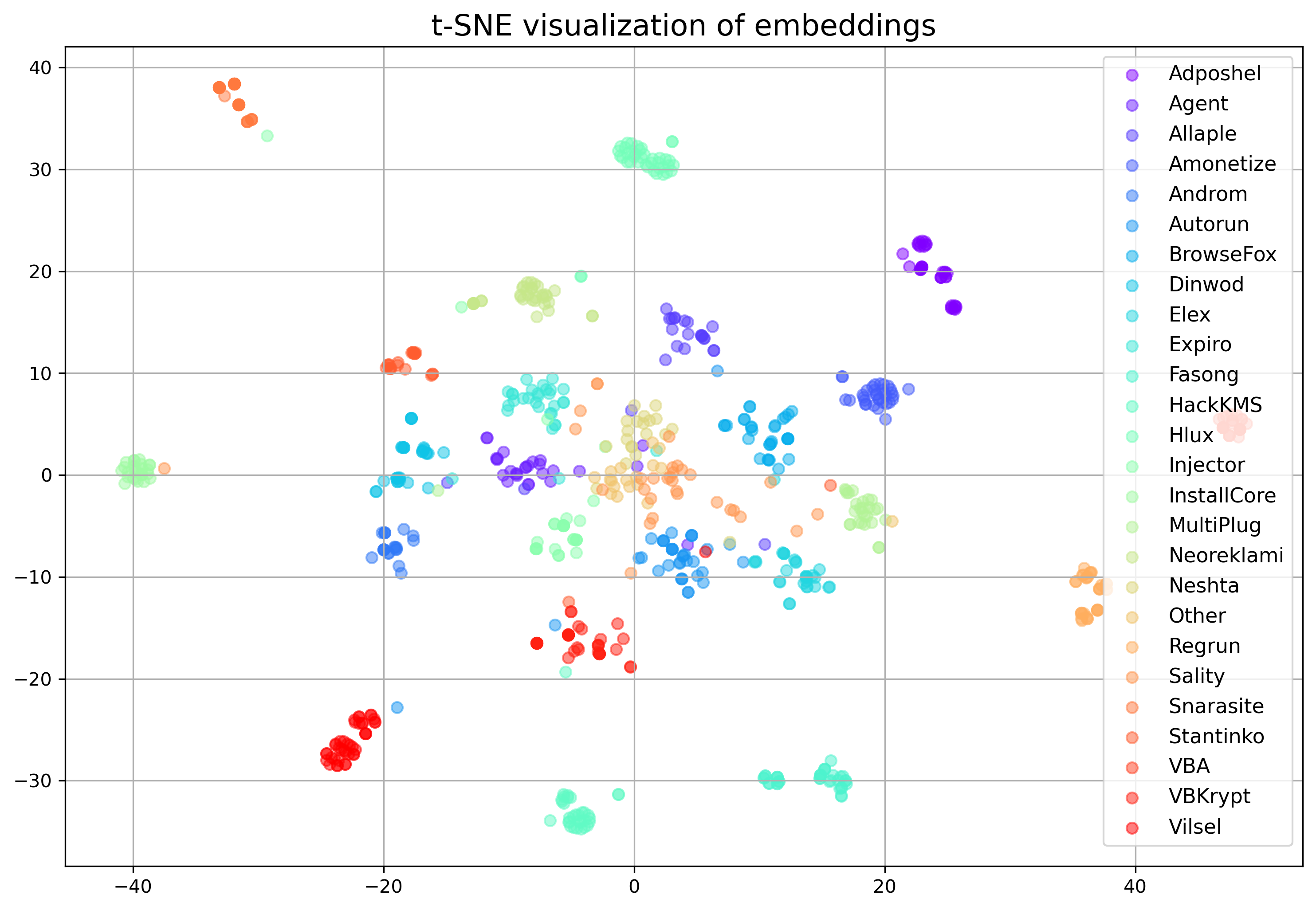} 
    \caption{Visualization of the embedding learned by the DIM-ProtoNetRF using the Malvi dataset.}
    \label{fig:m_emb}
\end{figure}

In Fig. \ref{fig:m_CM}, we provide the normalized confusion matrix to show the model's performance across the various classes. For each class, the diagonal elements show the percentage of correctly classified instances. Most classes have values closer to 1, indicating an outstanding classification performance.
 Fig. \ref{fig:m_emb} depicts the t-SNE visualization of the distribution of malware samples in a two-dimensional space. The colors show the 26 different classes in the MalVis dataset. The cluster of the same colors implies that the model's embedding effectively groups similar types of malware. 
This indicates that the proposed approach has learned meaningful representations from the dataset using the DIM model and building reliable prototypes. Moreover, it demonstrates the approach's capability to recognize commonalities and fundamental patterns within malware instances using the ProtoNet.

\begin{figure}[h]
    \centering
    \begin{minipage}{0.45\textwidth}
        \centering
        \includegraphics[width=\textwidth]{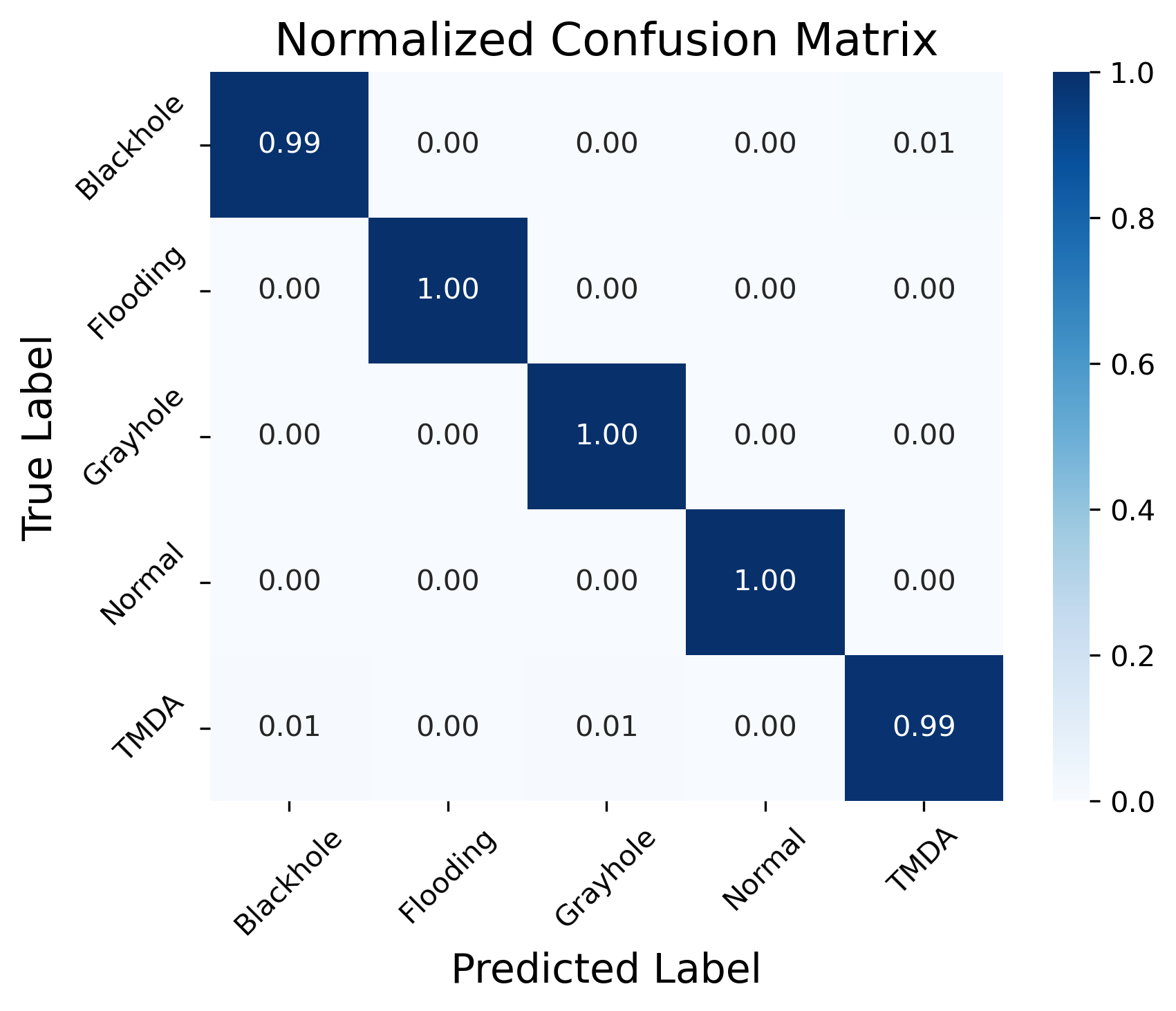} 
        \caption{Normalized confusion matrix of the WSN-DS dataset}
        \label{fig:W_CM}
    \end{minipage}\hfill
    \begin{minipage}{0.45\textwidth}
        \centering
        \includegraphics[width=\textwidth]{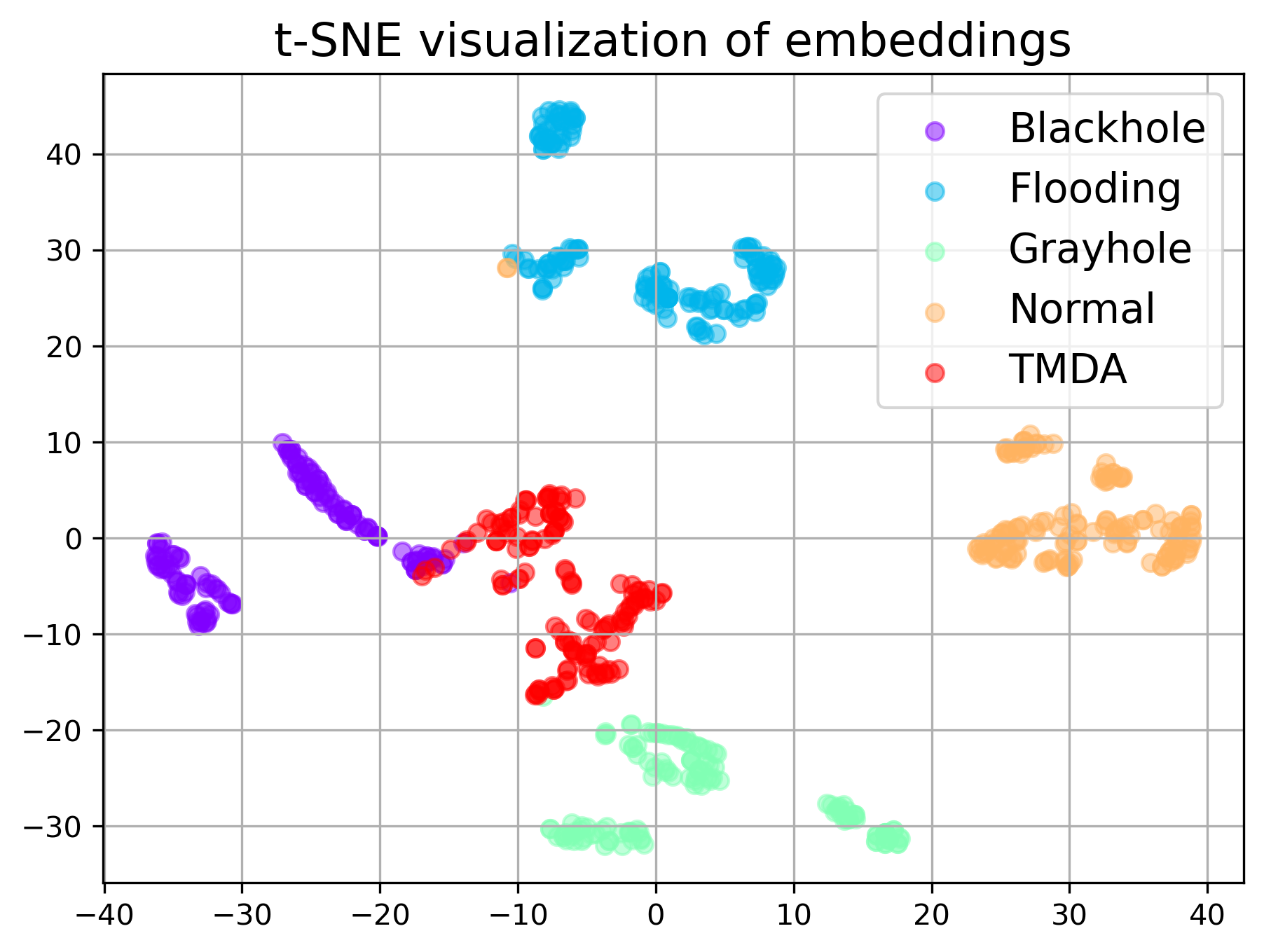} 
        \caption{Visualization of the embedding learned by the DIM-ProtoNetRF using the WSN-DS dataset}
        \label{fig:w_emb}
    \end{minipage}
\end{figure}

To validate the generalizability of the proposed approach, we followed the same experiments using the visualized WSN-DS dataset.
In Table \ref{tab:d2}, the intrusion detection and classification results are provided. 
The proposed DIM-ProtoNetRF performs well in intrusion detection, achieving an accuracy of 99.56 \%.
This good performance is also demonstrated in  Fig. \ref{fig:W_CM}, which shows the normalized confusion matrix. Besides, the embedding visualization presented in  Fig. \ref{fig:w_emb} shows a clear grouping of similar samples. These results highlight our method's stability and efficiency across various datasets.
\vspace{-0.3cm}
\begin{table}[h]
\centering
\caption{Performance results of the proposed RF-based DIM-Protonet using the WSN-DS dataset }
\label{tab:d2}
\begin{tabular}{|c|ccc|}
\hline
\textbf{Class} & \multicolumn{1}{c|}{\textbf{Precision}}           & \multicolumn{1}{c|}{\textbf{Recall}}                & \multicolumn{1}{c|}{\textbf{F1-score}} \\ \hline
Blackhole      & \multicolumn{1}{c|}{99.21}                          & \multicolumn{1}{c|}{99.21}                         &  \multicolumn{1}{c|}{99.21}   \\ \hline
Flooding       & \multicolumn{1}{c|}{100}                         & \multicolumn{1}{c|}{100}                         & \multicolumn{1}{c|}{100}   \\ \hline
Grayhole       &\multicolumn{1}{c|}{99.21 }                        & \multicolumn{1}{c|}{100}                         & \multicolumn{1}{c|}{99.60}            \\ \hline
Normal         & \multicolumn{1}{c|}{100}                          & \multicolumn{1}{c|}{100}                         & \multicolumn{1}{c|}{100 }           \\ \hline
TMDA           &\multicolumn{1}{c|}{99.27}                          & \multicolumn{1}{c|}{98.55}                         & \multicolumn{1}{c|}{98.91}            \\ \hline
\rowcolor[HTML]{EFEFEF} 
accuracy       & \multicolumn{3}{c|}{\cellcolor[HTML]{EFEFEF}99.56}                                                                         \\ \hline

\end{tabular}
\end{table}

\subsection{Ablation Study}

We carried out an ablation study to analyze the contributions of individual parts within our proposed approach. This study includes removing critical components of the strategy and evaluating its impact on the overall performance. Firstly, we assessed the model efficacy after excluding the DIM model, which is essential to feature extraction. In addition, we evaluated the performance in the absence of the final RF classifier. The results of these tests are illustrated in Tables \ref{tab:table3} and \ref{tab:table4}, which shed light on the importance of each phase of the proposed approach.
To conclude, the DIM model is our approach's backbone for feature extraction. Eliminating it reduces the model's ability to capture meaningful representations from the data, leading to a notable decrease in performance. In addition, the final RF classifier plays a vital role in improved decision-making. Its absence results in the loss of the classification capability, significantly impacting the model's overall performance.
\vspace{-0.3cm}
\begin{table}[h]
\caption{Results of the ablation study using MaleVis dataset}
\label{tab:table3}
\resizebox{0.99\textwidth}{!}{%
\begin{tabular}{|c|ccccccccc|}
\hline
\textbf{Model}     & \multicolumn{3}{c|}{\begin{tabular}[c]{@{}l@{}}\textbf{Classification} \\ \textbf{based on embedding}\end{tabular}}                                                        & \multicolumn{3}{c|}{\textbf{ ProtoNet}}                                                                           & \multicolumn{3}{c|}{ \begin{tabular}[c]{@{}l@{}} \textbf{Proposed }\\ \textbf{DIM-ProtoNetRF} \\ \textbf{ classifier} \end{tabular}}                                  \\ \hline
\textbf{Metric}     & \multicolumn{1}{c|}{\textit{Precision}} & \multicolumn{1}{c|}{\textit{Recall}} & \multicolumn{1}{c|}{\textit{F1-score}} & \multicolumn{1}{c|}{\textit{Precision}} & \multicolumn{1}{c|}{\textit{Recall}} & \multicolumn{1}{c|}{\textit{F1-score}} & \multicolumn{1}{c|}{\textit{Precision}} & \multicolumn{1}{c|}{\textit{Recall}} & \textit{F1-score} \\ \hline
\textbf{Macro avg} & \multicolumn{1}{c|}{95.55}              & \multicolumn{1}{c|}{95.51}           & \multicolumn{1}{c|}{95.51}             & \multicolumn{1}{c|}{94}                 & \multicolumn{1}{c|}{94}              & \multicolumn{1}{c|}{94}                & \multicolumn{1}{c|}{98.34}              & \multicolumn{1}{c|}{98.55}           & 98.36             \\ \hline
\textbf{Weighted}  & \multicolumn{1}{c|}{95.54}              & \multicolumn{1}{c|}{95.48}           & \multicolumn{1}{c|}{95.49}             & \multicolumn{1}{c|}{95}                 & \multicolumn{1}{c|}{94}              & \multicolumn{1}{c|}{94}                & \multicolumn{1}{c|}{98.79}              & \multicolumn{1}{c|}{98.60}           & \multicolumn{1}{c|}{98.63}             \\ \hline
\textbf{Accuracy}  & \multicolumn{3}{c|}{95.48}                                                                                              & \multicolumn{3}{c|}{94.35}                                                                                              & \multicolumn{3}{c|}{98.60}                                                                         \\ \hline
\end{tabular}}
\end{table}

 
\begin{table}[]
\caption{Results of the ablation study using WSN-DS dataset}
\label{tab:table4}
\resizebox{0.99\textwidth}{!}{%
\begin{tabular}{|c|ccccccccc|}
\hline
\textbf{Model}     & \multicolumn{3}{c|}{\begin{tabular}[c]{@{}l@{}}\textbf{Classification} \\ \textbf{based on embedding}\end{tabular}}                                                        & \multicolumn{3}{c|}{\textbf{ ProtoNet}}                                                                           & \multicolumn{3}{c|}{ \begin{tabular}[c]{@{}l@{}} \textbf{Proposed }\\ \textbf{DIM-ProtoNetRF} \\ \textbf{ classifier} \end{tabular}}                                  \\ \hline
\textbf{Metric}     & \multicolumn{1}{c|}{\textit{Precision}} & \multicolumn{1}{c|}{\textit{Recall}} & \multicolumn{1}{c|}{\textit{F1-score}} & \multicolumn{1}{c|}{\textit{Precision}} & \multicolumn{1}{c|}{\textit{Recall}} & \multicolumn{1}{c|}{\textit{F1-score}} & \multicolumn{1}{c|}{\textit{Precision}} & \multicolumn{1}{c|}{\textit{Recall}} & \textit{F1-score} \\ \hline
\textbf{Macro avg} & \multicolumn{1}{c|}{98.18}              & \multicolumn{1}{c|}{98.16}           & \multicolumn{1}{c|}{98.17}             & \multicolumn{1}{c|}{97.31}                 & \multicolumn{1}{c|}{97.27}              & \multicolumn{1}{c|}{97.28}                & \multicolumn{1}{c|}{99.54}              & \multicolumn{1}{c|}{99.54}           &    99.54          \\ \hline
\textbf{Weighted}  &  \multicolumn{1}{c|}{98.18}                 & \multicolumn{1}{c|}{98.16}              & \multicolumn{1}{c|}{98.17}                & \multicolumn{1}{c|}{97.42}              & \multicolumn{1}{c|}{97.36}           & \multicolumn{1}{c|}{97.38}     &   \multicolumn{1}{c|}{99.56}              & \multicolumn{1}{c|}{99.56}           & \multicolumn{1}{c|}{99.56}                 \\ \hline
\textbf{Accuracy}  & \multicolumn{3}{c|}{98.16}                                                                                              & \multicolumn{3}{c|}{97.36}                                                                                              & \multicolumn{3}{c|}{99.56}                                                                         \\ \hline
\end{tabular}}
\end{table}
 
\subsection{Discussion}
IoT networks are particularly vulnerable because of their extensive scale and connected devices' heterogeneity. 
In this work, combining SSL, FSL and RF techniques has significantly enhanced image-based malware detection in IoT environments. SSL has empowered the proposed approach to learn rich feature representations from the vast amounts of unlabeled data collected from IoT setups. Furthermore, FSL addresses the class imbalance issue that frequently arises in identifying uncommon and dangerous malware classes by enabling learning from a small number of labeled samples.
The proposed DIM-ProtoNetRF demonstrated good performance, achieving an accuracy rate of 99.6\% on the WSN-DS dataset and 98.6\% on the MaleVis dataset. 
The techniques used enable the model to learn effectively from a limited set of labeled samples. This ability to adapt is essential, particularly for dynamic IoT environments, where collecting massive labeled data for each attack type is challenging. 
The discussed results highlight the ability of the proposed approach to deal with the complexity of intrusion detection within IoT networks. 
In addition, the dynamic nature of cybersecurity threats requires systems that can quickly adapt to new and evolving attacks. The flexibility demonstrated by our FSL component in adapting to new threats with minimal data points is a step in the right direction. Our approach shows potential for expanding to handle more extensive and varied datasets, which is essential for real-world IoT applications. 
\vspace{-0.3cm}
\section{Conclusion}
\vspace{-0.3cm}
In this paper, we presented a novel approach called DIM-ProtoNetRF for intrusion detection in IoT networks, which consists of three phases: (1) Using SSL through the DIM model to extract reliable representations from unlabelled data; (2) Employing FSL with a prototypical network to enable the model to learn effectively from a small set of labeled examples; and (3) Applying an RF classifier to detect and classify various classes of attacks accurately.
The proposed approach uses only 20\% of the labeled training samples, reducing the need for a large amount of labeled data. It achieved good performance through extensive experimental analysis of two IoT datasets.
For future research directions, we aim to validate the application of our approach in real-time settings, enhancing its capability to detect and respond to threats instantly. Furthermore, we seek to test and refine the model across different IoT domains and configurations, ensuring its effectiveness across a broader range of network environments and attack vectors. 
\vspace{-0.3cm}

\vspace{-0.3cm}
%
%

\bibliographystyle{splncs04}  
\bibliography{references}  
\end{document}